\documentstyle[twocolumn,prl,aps,epsfig]{revtex}
\tighten
\newcommand{\beq}{\begin{eqnarray}}
\newcommand{\eeq}{\end{eqnarray}}
\renewcommand{\vec}[1]{{\mathbf{#1}}}
\begin{document}
\draft
\input epsf.sty

\title
{Nonlinear Transport Near a Quantum Phase Transition in Two Dimensions}
\author{Denis Dalidovich$^a$ and Philip Phillips$^b$}
\vspace{.05in}

%
\address
{$^a$National High Field Magnetic Laboratory,Florida State University,
Tallahassee, Florida\\
$^b$Loomis Laboratory of Physics,University of Illinois at Urbana-Champaign,
1100 W.Green St., Urbana, IL, 61801-3080}

%
\address{\mbox{ }}
\address{\parbox{14.5cm}{\rm \mbox{ }\mbox{ }
The problem of non-linear transport near a quantum phase transition is solved
within the Landau theory for the dissipative insulator-superconductor
phase transition in two dimensions.  Using the non-equilibrium Schwinger 
round-trip Green function formalism, we obtain
the scaling function for the non-linear conductivity in the quantum 
disordered regime. We find that 
the conductivity scales as $E^2$ at low field but crosses over at 
large fields to a universal constant on the order of $e^2/h$.  The crossover between these two regimes obtains when the length scale for the quantum fluctuations becomes comparable to that of the electric field within logarithmic accuracy.
}}
\address{\mbox{ }}
\address{\mbox{ }}

\maketitle

When an electric field is applied to a system near a quantum critical point,
non-linear transport obtains.
The non-linearity arises because the length scale\cite{fisher} 
associated with the electric field scales as $\ell_E\propto E^{-1/(1+z)}$, 
where $z$ is the dynamical exponent.  As it is the ratio of the electric 
field length scale to the correlation length, $\xi\approx\Delta^{-\nu}$, that 
enters the resultant DC resistivity, 
\beq\label{scaling}
\rho(T,E)=f\left(\Delta/T^{1/\nu z},\Delta/E^{1/\nu(z+1)}\right),
\eeq
a non-linear electrical response is inescapable. Here $\Delta$ is the 
distance from the critical point. Non-linear transport 
according to Eq. (\ref{scaling}) is expected to hold as long as the 
temperature is low enough so that the
length scale associated\cite{sondhi} with temperature, 
$\ell_T\approx 1/T^{1/z}$, exceeds that of the electric field,
$1/T^{1/z}\gg \ell_E$.   Because of the additional factor of $z$ that 
enters the electric field scaling contribution to the conductivity, 
simultaneous scaling of the resistivity with respect to temperature 
and electric field enables a direct determination of both the correlation 
and dynamical exponents, $\nu$ and $z$, respectively and hence a 
complete characterization of the critical properties.  Despite its 
obvious importance, electric field scaling persists as the outstanding 
problem in quantum criticality because {\bf no} theoretical account has 
been put forth to explain how quantum fluctuations conspire to yield 
non-linear transport.  The primary theoretical hurdle is simple: 
A successful theory of non-linear transport must lie outside the 
standard Kubo/linear response formalism.  
Experimentally, the problem is complicated by the fact that in a 
wide range of systems exhibiting quantum critical 
points\cite{yazdani,rimberg,shahar,kravchenko}, 
the  $I-V$ characteristics which are highly non-linear at small fields 
all become {\it linear}
at large fields.  In fact, the $I-V$ characteristics for varying values 
of $|\Delta|$ all attain a universal slope as $E\rightarrow \infty$. 
The vanishing of the non-linearity at high values of $E$ implies that 
the scaling function for the resistivity is a highly non-trivial and 
non-monotonic function of the electric field.   Hence, a successful theory 
of non-linear transport must uncloak how such non-monotonicity arises.

Using a non-equilibrium formalism, we calculate explicitly the scaling 
function for the non-linear conductivity for both the large and small 
electric field limits.  While the formalism can be applied to any 
critical theory, we focus on the Landau theory for the 
insulator-superconductor transition in which the dynamics are 
determined by an Ohmic bath; that is, $z=2$.  Within the 
Schwinger\cite{schwinger} round-trip double-time Green function formalism (also known
as the Kadanoff-Baym\cite{kb}, Keldysh formalism\cite{keldysh}), 
we evaluate the current for arbitrary electric field simply by 
calculating the appropriate Green function. When temperature is the smallest
parameter (quantum disordered regime), a crucial ratio which enters both
the conductivity and the inverse correlation length, $m^2$, is the ratio
\beq 
Q=(e^*E)^{2/3}/\Delta
\eeq
($e^* =2e$- the charge of a Cooper pair).  
For $Q\ll 1$, corrections due to field are sub-dominant and
$m^2=\Delta/\ln(1/\Delta))$ whereas for large field, 
$m^2= \Upsilon (e^*E)^{2/3}/\ln(1/(e^*E)^{2/3}$, $\Upsilon$ a constant.  
In the quantum disordered regime, the static conductivity scales 
as $Q^2$ for $Q\ll 1$, whereas in the large-field limit,
$Q\gg 1$, linear transport emerges with
a universal constant, $\sigma(E)=0.46 e^2/h$, that is in qualitative 
agreement with experiments\cite{yazdani,rimberg,shahar,kravchenko}.

The starting point for our analysis is the minimal Ginsburg-Landau action
(in the imaginary time)  
\beq\label{gaussian}
&&F[\psi]=\int d^2r\int d\tau\left\{
|\left(\nabla-\frac{ie^*}{\hbar}\vec A(\vec r,\tau)\right)
\psi(\vec r,\tau)|^2+\right.\nonumber\\
&&\left.\left|\partial_\tau\psi(\vec r,\tau)\right|^2
+\delta\left|\psi(\vec r,\tau)\right|^2 
+(U/2)\left|\psi(\vec r,\tau)\right|^4
\right\}+L_{\rm dis} 
\eeq
required to model quantum fluctuations and dissipation near the 
zero-resistance quantum critical point.
In Eq. (\ref{gaussian}), $\vec A(\vec r,\tau)$ 
is the vector potential, $e^*=2e$, $\delta$ is the bare distance to the 
quantum critical point. In terms of the Matsubara frequencies, 
the dissipation term, 
$L_{\rm dis}=\eta\sum_{\vec k,\omega_n}|\omega_n||\psi(\vec k,\omega_n)|^2$,
corresponds to the phenomenological Ohmic model introduced by 
Caldeira and Leggett~\cite{cl}, in which $\eta$ measures the strength of 
the dissipation. The order parameter 
$\psi(\vec r,\tau)$ is the standard two-component complex field 
whose expectation value is non-zero
in the superconducting phase. Our motivation for introducing 
dissipation, that is expected to be strong ($\eta\sim 1$),
is two-fold. First, the second-derivative term now becomes irrelevant. 
Consequently, the zero temperature transition falls into the $z=2$ 
universality class. Such a universality class describes 
a 2D insulator to high-$T_{c}$ superconductor transition in which 
superconductivity is destroyed by impurities\cite{herbut}.
Second, as we will see, $z=2$ dynamics are inherently easier to solve using 
the Schwinger\cite{keldysh,lp} technique than is the corresponding $z=1$ 
problem. Nonetheless, the generality of our conclusions leads us to 
believe that similar results must hold for the $z=1$ case as well.

As the problem we wish to treat is inherently out of equilibrium, we 
resort to the real-time `round-trip' Schwinger 
formalism\cite{schwinger,kb,keldysh,lp} which is ideally suited for 
solving problems in which an asymmetry exists between forward and 
backward evolution in time. We will be brief in our presentation of 
this technique as it is well-documented in the 
literature\cite{keldysh,lp,wilkins}
and our notation will follow that of Lifshitz and Pitaevskii\cite{lp}. 
To use the real-time formalism, one needs to rewrite the action, 
Eq. (\ref{gaussian},  as an integral over the Keldysh contour.
A constant in space and time electric field is assumed to enter 
the action via the vector potential, $\vec A(t) =-\vec E t$.
This gauge choice allows for all functions to be Fourier expandable
in space, since full translational invariance is retained.
We define the time ordered and anti-time ordered Green functions    
\beq\label{gf1}
iG^{--}_{\vec p}(t_1,t_2)&=&FT
\langle T[\psi(\vec r,t_1),\psi^{*}(\vec r',t_2)]\rangle\nonumber\\
iG^{++}_{\vec p}(t_1,t_2)&=&FT\langle[{\tilde T}\psi(\vec r,t_1),
\psi^{*}(\vec r',t_2)]\rangle,
\eeq
where $T$ and $\tilde T$ are the time ordering and inverse time
ordering operators.
Here $FT$ represents the Fourier transform with respect to space.
We will need also the non-time ordered functions 
\beq\label{gf2}
iG^{-+}_{\vec p}(t_1,t_2)&=&FT\langle \psi^{*}(\vec r',t_2)
\psi(\vec r,t_1)\rangle\nonumber\\
iG^{+-}_{\vec p}(t_1,t_2)&=&FT\langle \psi(\vec r,t_1)
\psi^{*}(\vec r',t_2)\rangle,
\eeq
the last of which is directly related to the expectation value of 
the current operator through
\beq
\vec J(t)=\frac{-2ie^\ast}{\hbar}\int \vec p \frac{d^d p}{(2\pi)^d}
G^{-+}_{\vec p}(t,t).
\eeq
The average values in these expressions are over all states of the system, 
not simply equilibrium ones. Since the four functions in Eqs. (\ref{gf1})
and (\ref{gf2}) forming a 2 by 2 matrix ${\hat G}$, are not independent,
one employs the so-called Keldysh rotation to work with the three independent 
functions $G^{R}(G^{A})=G^{--}-G^{-+}(G^{+-})$,$G^{K}=G^{-+}+G^{+-}$.
The matrix of the exact Green functions ${\hat G}$
is connected to that of the Green functions for free quasi-particles 
${\hat G^{(0)}}$ via the Dyson equation\cite{lp},
${\hat G}= {\hat G^{(0)}}+\int\int {\hat G^{(0)}}{\hat \Sigma}{\hat G}$, 
in which the integration over internal time and space arguments is assumed 
and the self-energy matrix ${\hat \Sigma}$ is itself in general 
a complicated functional of ${\hat G}$. 
For the static field, the resulting Green functions $G^{+-},G^{-+},G^{K}$, 
taken at $t=t_1=t_2$ do not depend on $t$, reflecting the 
fact that the system is time translationally invariant as well.

The non-linear conductivity is defined as the constant of proportionality 
between the current and the electric field. We will orient the field along 
x-axis and hence $ J_{xx}(E)=\sigma_{xx}(E)E$. 
In the Schwinger formalism, the direct relationship\cite{wilkins} 
between $G^{-+}$ and the retarded and advanced Green functions 
\beq\label{G-+}
G^{-+}_{\vec p}(t_1,t_2)=-\int dt_3dt_4G^{R}_{\vec p}(t_1,t_3)
\Sigma^{-+}(t_3,t_4)
G^{A}_{\vec p}(t_4,t_2)
\eeq 
is obtained by formally solving the Dyson equation.
In Eq. (\ref{G-+}), note the minus sign that is absent in the fermion 
problem\cite{wilkins}. Analogous expressions hold for $G^{+-}$ ($G^{K}$) 
with $\Sigma^{-+}$ replaced with $\Sigma^{+-}$ ($\Sigma^{K}$).
In general, the self-energy ${\hat \Sigma}$ arises from the interaction $U$,
${\hat \Sigma_{U}}$, and the coupling to the dissipative bath 
${\hat \Sigma_{d}}$.
However, treating the interactions in the large-N (mean-field) limit,
we obtain that this approximation gives rise only to the
renormalization of the bare distance to the critical point in 
the action\cite{chamon},
\beq\label{m2}
m^2=\delta-\frac{U}{2}\int\frac{d^2 p}{(2\pi)^2}iG^K(p,t,t),
\eeq
so that a new Gaussian action with the excitation spectrum 
$\epsilon(\vec p)=\vec p^2+m^2$ obtains. 
There is no contribution from interactions to $\Sigma^{-+}$ 
(and $\Sigma^{K}$) at this level, and the corresponding Green functions
are assessed by substituting $\Sigma_{d}(t_3 -t_4)$ into Eq. (\ref{G-+}).
To this end, we simplify the notation by setting $\eta$ equal to unity, 
so that in the frequency space\cite{chamon}
$\Sigma_{d}^{-+}(\omega)=-2i\omega e^{-|\omega|/\Lambda}n_B(\omega)$,
$\Sigma_{d}^{R}(\omega)=\Sigma^{A}(\omega)^*=i\omega e^{-|\omega|/\Lambda}$,
$\Sigma_{d}^{K}(\omega)=2i\omega e^{-|\omega|/\Lambda}(2n_B(\omega)+1)$,
with $n_B(\omega)$ the distribution function for bosons. The upper 
frequency cutoff $\Lambda$ is necessary only to ensure the convergence of 
the zero temperature parts of the Fourier integrals 
$\Sigma_{d} (t)=\int \Sigma_{d} (\omega)(d\omega /2\pi)e^{-i\omega t}$.

The equation of motion for the corresponding retarded Green function
\beq
\frac{\partial}{\partial t_1}G^R_{\vec p}(t_1,t_2)+
\epsilon(\vec p-e^* \vec E t_1)G^R_{\vec p}(t_1,t_2)=\delta(t_1-t_2)
\eeq 
has a simple solution
\beq\label{GR}
G^R_{\vec p}(t_1,t_2)=\theta(t_1-t_2)\exp
\left\{-\int_{t_1}^{t_2}\epsilon(\vec p-e^* \vec E \tau)d\tau\right\}.
\eeq
In $G^A$, the order of the time arguments is reversed
in the step function $\theta(x)$.  Consequently, after Eqs. (\ref{G-+}), 
and (\ref{GR}) are combined, our problem is, in principle, 
solved. To make contact with the case $E=0$, 
we note that in the absence of a field, the frequency Fourier transform 
of Eq. (\ref{GR}) reduces to the expected result 
$(\epsilon(\vec p)-i\eta\omega)^{-1}$.

When the field is non-zero, two distinct regimes emerge.  To obtain an estimate
of the crossover field, we determine at which value of the field can 
the exponential in $G^R$ be expanded. We rewrite 
$\epsilon(\vec p)$ in the exponential 
of Eq. (\ref{GR}) as $p^2+2p_x e^*E \tau+m^2$.  Upon integrating 
the resultant exponential 
in Eq. (\ref{GR}), we find that the electric-field dependence can be 
removed from the exponential provided that $p_x  e^*E t^2 \ll 1$.  
Since both momenta and time are integrated over, this criterion is never 
satisfied. Hence, strictly speaking, the electric field effects are always 
non-perturbative.   Nonetheless, if we focus on the characteristic 
timescale, $t\propto 1/m^2$ and momentum scale, $p_x\propto m$, 
we find that 
$e^*E /m^3<< 1$ or equivalently, $(e^*E)^{2/3}/m^2=Q\ll 1$ 
determines the small and large field regimes.  It is this criterion, 
rather than $1/T^{1/z}\gg \ell_E$, which will serve to determine when 
non-linear transport according to Eq.(\ref{scaling}) holds. 

To obtain the conductivity, we need first calculate $G^{-+}$. We 
 perform the integration in the exponent of Eq. (\ref{GR}) and switch to 
the Wigner coordinates\cite{wilkins,lp}: $t=t_1-t_2$ and $u=(t_1+t_2)/2$.  
The Green function from which the current is obtained reduces to
\beq\label{Gmpm}
-iG^{-+}_{\vec p}(t,t)=\int_0^\infty du\int_{-2u}^{2u} dt 
M(\vec p,e^* E;,u,t)K(t)
\eeq
where $K(t)=-i\Sigma^{-+}(-t)$ and 
\beq
M(\vec p,e^* E;u,t)&=&\exp\left\{-\left(2\epsilon(\vec p)u-2p_x
e^* E(2u^2+t^2/2)\right.\right.\nonumber\\
&&\left.\left.+\frac{(e^* E)^2}{3}(u^3+3ut^2/2)\right)\right\},
\eeq
and the kernel $K(t)$ is obtained as a result of the appropriate integration 
over $\omega$ with the cutoff $\Lambda$ necessary to regularize 
the $T=0$ part. The result is
\beq\label{k1}
K(t)&=&\frac{1}{\pi}\frac{\Lambda^{-2}-t^2}{(\Lambda^{-2}+t^2)^2}+\frac{1}{\pi}
\left[\frac{1}{t^2}-\frac{\pi^2 T^2}{\sinh^2(\pi Tt)}\right]\nonumber\\
&=&K_0(t)+K_T (t)
\eeq
in which the second temperature dependent part $K_T (t) \approx \pi T^2 /3$
as $Tt \ll 1$. 
Similarly, because $G^K=G^{-+}+G^{+-}$, we can also express the inverse 
correlation length
\beq\label{m2new}
m^2=\delta+U\int_{\vec p}\int_0^\infty du\int_{-2u}^{2u}dt
M(\vec p,e^* E;u,t)K(t)
\eeq
in terms of the same functions appearing in $G^{-+}$. 

Our problem has been reduced then to the computation of two types of coupled 
integrals in Eqs. (\ref{m2new}) and (\ref{Gmpm}). In the $T\rightarrow 0$
limit, the main difficulty is to obtain the cutoff-free expressions for $m$ 
and $J$, since the physical results should obey the universal scaling
and correspond to $\Lambda^{-1}=0$.  
In this regard, we have found the following procedure most helpful: 
1) Regularize the zero-temperature part 
(that is, the integrals over the product $M(\vec p,e^* E;u,t)K_0(t)$) 
by adding and subtracting $M(\vec p,e^* E;u,0)$ and $M(\vec p,0;u,0)$ 
to $M(\vec p,e^* E;u,t)K_0(t)$. In the current, the integral 
over $\vec p$ in the part involving $M(\vec p,0;u,0)$ vanishes identically. 
In Eq. (\ref{Gmpm}) similar contribution gives rise to the frequency 
and momentum dependent renormalizations of 
$\delta$, as well as the term $m^2 \ln (1/m^2)$, ensuring the validity
of the subsequently used logarithmic accuracy.  
2) In the part containing 
$M(\vec p,e^* E;u,0)-M(\vec p,0;u,0)$, perform straightforwardly 
the integration over $t$ in $K_0(t)$ to yield $2/(u\pi)$. 
3) In the part containing 
$M(\vec p,e^* E;u,t)-M(\vec p,e^* E ;u,0)$, 
 set $\Lambda^{-1}=0$ as this term is completely convergent. 
4) In the terms containing the afore-mentioned 
differences, we introduce the change of variables 
$p_x\rightarrow p_x-(e^* E) u/2(1+t^2/4u^2)$, and perform the Gaussian
integration over momenta $p_x$ and $p_y$. Similar integration is
performed in the part containing $K_{T}$. 
5) Finally, change variables to $y=t/2u$ and $z=u(e^* E)^{2/3}$. 
These transformations lead to the appearance of the auxilliary functions 
\beq
&&M_2(z)=\exp \{-\frac{2m^2z}{(e^* E)^{2/3}} \},\nonumber\\
&&M_1(z,y)= M_2(z)\exp\{-z^3(1/6+y^2-y^4/2)\},
\eeq
and $\tilde{K}_{T}(z,y)=K_T(t=2z(e^* E)^{4/3}y)$.
The expressions determining the current reduces now 
to quadrature:
\beq\label{jx}
J&=&\frac{2e^* (e^* E_)}{16\pi^2\hbar}\left\{\int_0^\infty\frac{dz}
{z}\int_{0}^{1}\frac{dy}{y^2}f(y)\left[M_1(z,0)-M_1(z,y)
\right]\right.\nonumber\\
&&\left.+\frac{4\pi}{(e^* E)^{4/3}}\int_0^\infty z dz\int_{0}^{1}dy 
\tilde{K}_T(z,y)f(y)M_1(z,y)\right\}
\eeq
with $f(y)=(1+y^2)$. Similarly, within logarithmic accuracy, we have
\beq\label{msq}
&&m^2 \ln (1/m^2)=\Delta +(e^* E)^{2/3}\left\{ \int_0^\infty\frac{dz}
{2z^2}\int_{0}^{1}\frac{dy}{y^2} \left[ M_1(z,0)\right.\right.
\nonumber\\
&&\left.\left. -M_1(z,y)\right]
-\int_0^\infty\frac{dz}{2z^2}\left[ M_2 (z)-M_1(z,0)
\right]\right.\nonumber\\
&&\left.+\frac{2\pi}{(e^* E)^{4/3}} \int_0^\infty dz \int_0^{1}dy
\tilde{K}_T(z,y)M_1(z,y)\right\},
\eeq
where $\Delta$ is the renormalized distance to the quantum critical point.

Consider now the quantum-disordered regime, in which $T$ is the smallest parameter. At low electric fields, $Q=(e^* E)^{2/3}/\Delta\ll 1$, we expand  Eqs. (\ref{msq}) and (\ref{jx}) in powers of $Q$ to obtain
\beq\label{m2ee}
m^2=\frac{\Delta}{\ln(1/\Delta)}+\frac{(e^* E)^2}
{12(\Delta^2/\ln(1/\Delta))}
+\frac{\pi^2}{3}\frac{T^2}{\Delta/\ln(1/\Delta)},
\eeq
and
\beq\label{Jsmall}
J=\frac{4e^2}{h}\left[\frac{\pi T^2}{9m^4}+\frac{(e^*E)^2)}
{15\pi m^6}\right] E =\sigma(E) E,\quad Q\ll 1.
\eeq
for the inverse correlation length and the non-linear current-voltage response.  The first term matches 
identically with the value of the linear current response calculated 
earlier\cite{dp} for the quantum disordered regime.  Most importantly, 
the current scales as the third power of the voltage in the quantum 
disordered regime. This result clearly could not have been obtained within 
linear response theory.  This non-linear response gives rise to a non-zero 
conductivity (albeit non-linear) even at $T=0$ on the putative insulating 
side of the transition.  Such a term arises entirely from the zero-temperature 
part of the integrals in Eq. (\ref{jx}).  As such a term will always be 
present regardless of which universality class describes the quantum phase 
transition, we predict that the $T=0$ non-linear conductivity will
always remain non-zero on the disordered side of the transition.  This 
conclusion is borne out experimentally by the series of measurements 
of $dI/dV$ in insulator-superconductor\cite{yazdani,rimberg}, quantum Hall 
to insulator\cite{shahar}, and insulator to metal 
transitions\cite{kravchenko}. Further, the form of the non-linearity, 
$E^3$, is non-trivial and is also in agreement with the pronounced 
curvature in the $I-V$ curves on the insulating side of the dissipative 
insulator-superconductor transition\cite{rimberg}.  To explore 
the large field regime from Eq. (\ref{jx}), we note that when $Q\gg 1$,
$e^* E$ is the only parameter that determines $m$. Anticipating the 
logarithmic smallness, we set $M_2(z)\approx 1$, and after performing 
first the integral over $z$, we obtain
\beq\label{m2e}
m^2=\frac{\Upsilon (e^* E)^{2/3}}{\ln(1/(e^* E)^{2/3})},
\eeq
where
\beq
\Upsilon&=&\frac{6^{2/3}}{12}\Gamma \left( \frac{2}{3} \right)\int_0^1
\left[ \frac{(1+6y^2-3y^4)^{1/3}- f(y)}{y^2} \right] dy \nonumber\\
&&=0.116525.
\eeq
Here, as before, $f(y)=1+y^2$.
In the same limit, the current reduces to a single integral of the form,
\beq\label{nonlinc}
J&=&\frac{2e^*}{4\pi^2 \hbar}\frac{e^* E}{12}\int_0^1\frac{dy}{y^2}f(y)
\ln(1+6y^2-3y^4)\nonumber\\
&&=0.46 \frac{e^2}{h}E\quad\quad\quad Q\gg 1
\eeq
which proves that $dI/dV$ approaches a universal constant, which 
of course depends on the universality class, as $E\rightarrow\infty$.  
At present, the most we can conclude is that the constant is of order 
$e^2/h$. This is expected to be of order $e^2/h$ as it should be on 
the order of the normal sheet resistance.  Hence, we have developed a 
formalism which is capable of describing the experimental crossover from 
the non-linear to the linear regime at high field.   
 
The crossover to the linear regime has a fundamental origin.  Because at mean-field $\nu=1/2$ and $z=2$, we can rewrite Eq. (\ref{m2e}) as $m^2\propto 1/( \ell_E^{1/\nu}\ln(\ell_E^{1/\nu}))$. Hence, at large fields, the correlation length is cutoff by the electric field length scale, to logarithmic accuracy. That is, the only length scale in the large field limit is $\ell_E$.   As it is the product $(m\ell_E)^{\nu}$ that enters the non-linear conductivity, the electric field dependence naturally drops out and the conductivity approaches a universal constant. 
In the opposite regime, the correlation length and the electric field length scale are distinct as Eq. (\ref{m2ee}) indicates.  The non-linearity arises entirely from the quantum fluctuations on the length scale $\xi$.   In the non-linear regime, it is tempting to rewrite the current, Eq. (\ref{Jsmall}), in terms
of an effective temperature, $T^2_{\rm eff}=T^2+0.06(e^*E/m)^2$. The current then simplifies to
\beq
J=\frac{4\pi e^*}{9 h}T^2_{\rm eff} E.
\eeq
For electrons localized in the band tails of semiconductors, the concept of an electric field-dependent effective temperature has been used extensively\cite{shlovskii,phillips}. Electrons moving against the electric field are accelerated on a length scale set by the localization length. In the quantum disordered regime, it is the correlation length that plays the role of the localization length. Consequently, it is not surprising that the bosonic excitations can be described by an effective E-field dependent effective temperature.

This work was partially funded by donors of the ACS petroleum research fund
and the NSF DMR-0305864 .

\end{document}